\documentclass{llncs}
\usepackage[numbers,sort&compress,sectionbib]{natbib}
\usepackage{amsmath,graphicx}
\usepackage{amsfonts}
\usepackage{acro}
\usepackage{tabularx}
\graphicspath{{Figure/}}
\usepackage{color}
\usepackage{multirow}
\usepackage{siunitx}
\usepackage{booktabs} % To thicken table lines
 \usepackage[font={small}]{caption, subfig}
\usepackage{tikz}

\usepackage{epstopdf}
\usepackage{ulem}

\newcolumntype{Y}{>{\centering\arraybackslash}X}
\newcommand{\ie}{\textit{i.e.},}

\newcommand{\Sec}{\S}
\newcommand{\Fig}{Fig.}

\AtBeginDocument{}

\DeclareAcronym{GTV}{
short=GTV,
long=gross tumor volume
}

\DeclareAcronym{CTV}{
short=CTV,
long=clinical target volume
}

\DeclareAcronym{RTCT}{
short=RTCT,
long=radiotherapy computed tomography
}
\DeclareAcronym{PET/CT}{
short=PET/CT,
long=positron emission tomography/computed tomography
}

\DeclareAcronym{PET}{
short=PET,
long=positron emission tomography
}

\DeclareAcronym{CT}{
short=CT,
long=computed tomography
}

\DeclareAcronym{PSNN}{
short=PSNN,
long=progressive semantically nested network
}

\DeclareAcronym{RT}{
short=RT,
long=radiotherapy
}
 
\DeclareAcronym{CNN}{
short=CNN,
long=convolutional neural network
}

\DeclareAcronym{ASD}{
short=$\text{ASD}_{GT}$,
long=average surface distance with respect to the ground truth contour
}

\DeclareAcronym{EF}{
short=EF,
long=early fusion
}

\DeclareAcronym{LF}{
short=LF,
long=late fusion
}

\DeclareAcronym{DSC}{
short=DSC,
long=Dice score
}

\DeclareAcronym{HD}{
short=HD,
long=Hausdorff distance
}

\title{Accurate Esophageal Gross Tumor Volume Segmentation in PET/CT using Two-Stream Chained 3D Deep Network Fusion}
\author{Dakai Jin\textsuperscript{1}  \and Dazhou Guo\textsuperscript{1} \and Tsung-Ying Ho\textsuperscript{2} \and Adam P. Harrison\textsuperscript{1} \and Jing Xiao\textsuperscript{3} \and Chen-kan Tseng\textsuperscript{2} \and Le Lu\textsuperscript{1}  }

\institute{\textsuperscript{1}PAII Inc., Bethesda, MD, USA \\ \textsuperscript{2}Chang Gung Memorial Hospital, Linkou, Taiwan, ROC \\ \textsuperscript{3}Ping An Technology, Shenzhen, China \\
}

\begin{document}
\setlength{\abovecaptionskip}{1ex}
\setlength{\belowcaptionskip}{1ex}
\setlength{\floatsep}{1ex}

\maketitle

\begin{abstract}
\Ac{GTV} segmentation is a critical step in esophageal cancer radiotherapy treatment planning. Inconsistencies across oncologists and prohibitive labor costs motivate automated approaches for this task. However, leading approaches are only applied to \ac{RTCT} images taken prior to treatment. This limits the performance as \ac{RTCT} suffers from low contrast between the esophagus, tumor, and surrounding tissues. In this paper, we aim to exploit both \ac{RTCT} and \ac{PET} imaging modalities to facilitate more accurate \ac{GTV} segmentation. By utilizing \ac{PET}, we emulate medical professionals who frequently delineate \ac{GTV} boundaries through observation of the \ac{RTCT} images obtained after prescribing radiotherapy and PET/CT images acquired earlier for cancer staging. To take advantage of both modalities, we present a two-stream chained segmentation approach that effectively fuses the CT and \ac{PET} modalities via early and late 3D deep-network-based fusion. Furthermore, to effect the fusion and segmentation we propose a simple yet effective \ac{PSNN} model that outperforms more complicated models. Extensive 5-fold cross-validation on $110$ esophageal cancer patients, the largest analysis to date, demonstrates that both the proposed two-stream chained segmentation pipeline and the \ac{PSNN} model can significantly improve the quantitative performance over the previous state-of-the-art work by $11\%$ in absolute \ac{DSC} (from $0.654\pm0.210$ to $0.764\pm0.134$) and, at the same time, reducing the Hausdorff distance from $129\pm 73 mm$ to $47 \pm 56 mm$. %These results indicate that our approach can provide impactful improvements for esophageal \ac{GTV} segmentation.
\end{abstract}

%\begin{keywords}
%Esophageal gross tumor volume, PET/CT, segmentation
%\end{keywords}

\acresetall
\acuse{PET, CT}

\section{Introduction}
\label{sec:intro}

Esophageal cancer ranks sixth in mortality amongst all cancers worldwide, accounting for $1$ in $20$ cancer deaths~\cite{bray2018global}. Because this disease is typically diagnosed at late stages, the primary treatment is a combination of chemotherapy and \ac{RT}. One of the most critical tasks in \ac{RT} treatment planning is delineating the \ac{GTV}, which serves as the basis for further contouring the clinical target volume~\cite{jin2019ctv}. Yet, manual segmentation consumes great amounts of time and effort from oncologists and is subject to inconsistencies~\cite{tai1998variability}. Thus, there is great impetus to develop effective tools for automated \ac{GTV} segmentation. 

\begin{figure}[t]
\centering
\includegraphics[width=0.98\columnwidth]{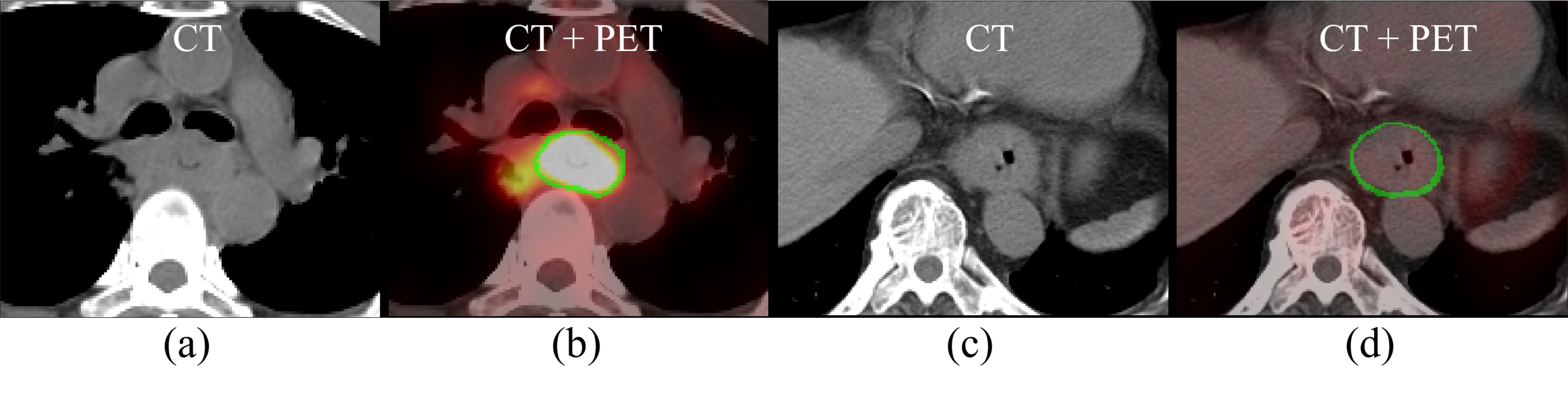}
\caption{Esophageal GTV examples in CT and PET images, where the green line indicates the \acs{GTV} boundary. (a)-(b): although the GTV boundaries are hardly distinguishable in CT, it can be reasonably inferred with the help of the PET image, in spite of other false positive high-uptake regions. (c)-(d) here, no high uptake regions appear in PET; however, the esophagus wall enlargement evident in CT may indicate the GTV boundary~\cite{iyer2004imaging}. }
\vspace{-2em}
\label{fig:demo}
\end{figure}

Deep \acp{CNN} have made remarkable progress in the field of medical image segmentation~\cite{Cicek2016,harrison2017progressive,jin20173d,jin2018ct,RothLLHFSS18}. Yet, only a handful of studies have addressed automated esophageal \ac{GTV} segmentation~\cite{Hao2017Esophagus, yousefi2018esophageal}, all of which rely on only the \ac{RTCT} images. The assessment of \ac{GTV} by \acs{CT} has been shown to be error prone, due to the low contrast between the \ac{GTV} and surrounding tissues~\cite{muijs2009consequences}. Within the clinic these shortfalls are often addressed by correlating with the patient's \ac{PET/CT} scan, when available. These \acp{PET/CT} are taken on an earlier occasion to help stage the cancer and decide treatment protocols. Despite misalignments between the \ac{PET/CT} and \ac{RTCT}, \acs{PET} still provide highly useful information to help manually delineate the \ac{GTV} on the \ac{RTCT}, due to its high contrast highlighting of malignant regions~\cite{leong2006prospective}.  As shown in Fig.~\ref{fig:demo}, \ac{CT} and \acs{PET} can each be crucial for accurate \ac{GTV} delineation, due to their complementary strengths and weaknesses.  While recent work has explored co-segmentation of tumors using PET and CT~\cite{zhong2018simultaneous,zhao2018tumor}, these works only consider the \ac{PET/CT} image. In contrast, leveraging the diagnostic \ac{PET} to help perform \ac{GTV} segmentation on an \ac{RTCT} image requires contending with the unavoidable misalignments between the two scans acquired at different times.

To address this gap, we propose a new approach, depicted in Fig.~\ref{fig:overall_pipeline}, that uses a two-stream chained pipeline to incorporate the joint \ac{RTCT} and \ac{PET} information for accurate esophageal \ac{GTV} segmentation. First, we manage the misalignment between the \ac{RTCT} and \ac{PET/CT} by registering them via an anatomy-based initialization. Next, we introduce a two-stream chained pipeline that combines and merges predictions from two independent sub-networks, one only trained using the \ac{RTCT} and one trained using both \ac{RTCT} and registered \ac{PET} images. The former exploits the anatomical contextual information in \ac{CT}, while the latter takes advantage of \ac{PET}'s sensitive, but sometimes spurious and overpoweringly strong contrast. The predictions of these two streams are then deeply fused together with the original \ac{RTCT} to provide a final robust GTV prediction. Furthermore, we introduce a simple yet surprisingly powerful \ac{PSNN} model, which incorporates the strengths of both UNet~\cite{Cicek2016} and P-HNN~\cite{harrison2017progressive} by using deep supervision to progressively propagate high-level semantic features to lower-level, but higher resolution features. Using 5-fold cross-validation, we evaluate the proposed approach on $110$ patients with \ac{RTCT} and \ac{PET}, which is more than two times larger than the previously largest reported dataset for esophageal \ac{GTV} segmentation~\cite{yousefi2018esophageal}. Experiments demonstrate that both our two-stream chained pipeline and the \ac{PSNN} each provide significant performance improvements, resulting in an average \ac{DSC} of $76.4\% \pm 13.4\%$, which is $11\%$ higher over the previous state-of-the-art method using DenseUNet~\cite{yousefi2018esophageal}.

\section{Methods}
\label{sec:method}

\begin{figure}
\centering
\includegraphics[width=1\columnwidth]{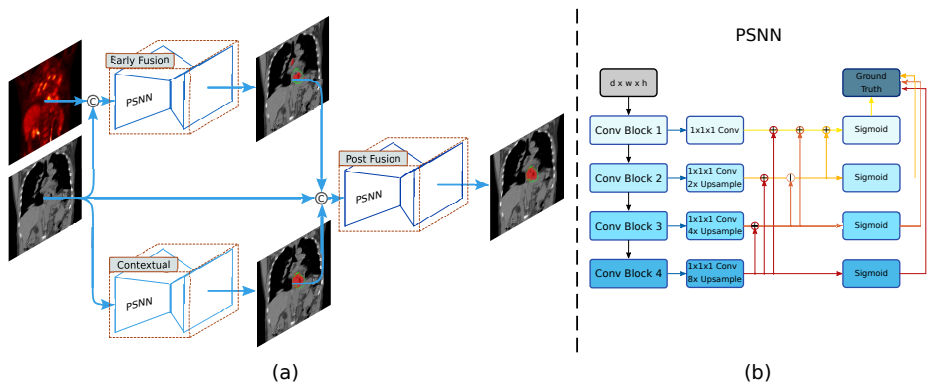}
\caption{(a) depicts our two-stream chained esophageal GTV segmentation method consisting of \acf{EF} and \acf{LF} networks, while (b) illustrates the \acs{PSNN} model, which employs deep supervision at different scales within a parameter-less high-to-low level image segmentation decoder. The first two and last two blocks are composed of two and three $3\times3\times3$ convolutional+BN+ReLU layers, respectively.}
\vspace{-2em}
\label{fig:overall_pipeline}
\end{figure}

\Fig~\ref{fig:overall_pipeline} depicts an overview of our proposed two-stream chained esophageal \ac{GTV} segmentation pipeline, which uses early and late 3D deep network fusions of \ac{CT} and \ac{PET} scans. Not shown is the registration step, which is detailed in \Sec\ref{sec:registration}. 

\subsection{PET to \ac{RTCT} Registration}
\label{sec:registration}

\begin{figure}
\centering
\includegraphics[width=0.96\columnwidth]{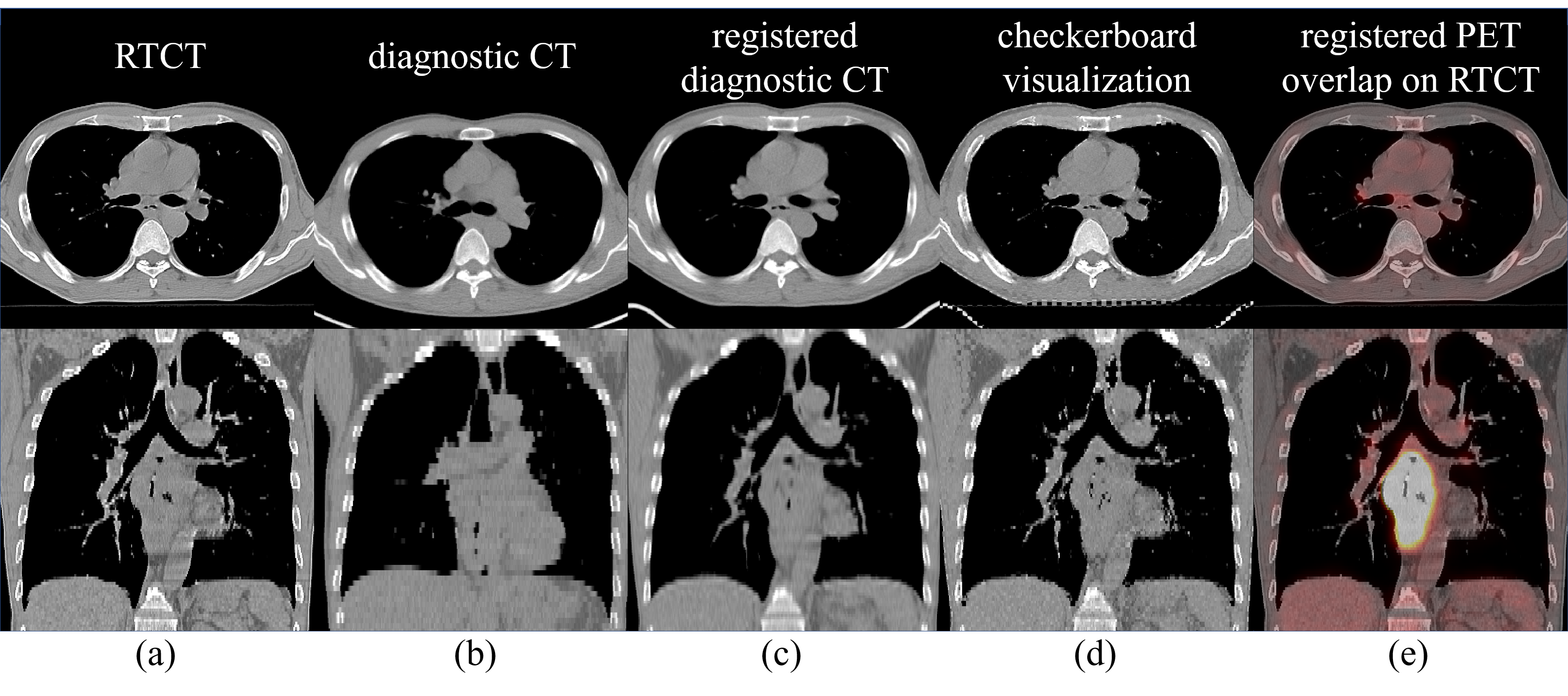}
\caption{Deformable registration results for a patient shown in axial and coronal views. (a) shows the RTCT image; (b, c) depicts the diagnostic CT image before and after the registration, respectively; (d) depicts a checkerboard visualization of the RTCT and registered diagnostic CT images; and (e) overlays the PET image, transformed using the diagnostic CT deformation field, on top of the RTCT.}
\vspace{-2em}
\label{fig:registration_demo}
\end{figure}

To generate aligned \ac{PET}/\ac{RTCT} pairs, we register the former to the latter. This is made possible by the diagnostic \ac{CT} accompanying the \ac{PET}. To do this, we apply the cubic B-spline based deformable registration algorithm in a coarse to fine multi-scale deformation process~\cite{rueckert1999nonrigid}. We choose this option due to its good capacity for shape modeling and efficiency in capturing local non-rigid motions. However, to perform well, the registration algorithm must have a robust rigid initialization to manage patient pose and respiratory differences in the two CT scans. To accomplish this, we use the lung mass centers from the two \ac{CT} scans as the initial matching positions. We compute mass centers from masks produced by the P-HNN model~\cite{harrison2017progressive}, which can robustly segment the lung field even in severely pathological cases. This leads to a reliable initial matching for the chest and upper abdominal regions, helping the success of the registration. The resulting deformation field is then applied to the diagnostic \ac{PET} to align it to the \ac{RTCT} at the planning stage. One registration example is illustrated in \Fig~\ref{fig:registration_demo}.

\subsection{Two-Stream Chained Deep Fusion}
As mentioned, we aim to effectively exploit the complementary information within the \ac{PET} and \ac{CT} imaging spaces. To do this, we design a two-stream chained 3D deep network fusion pipeline. Assuming $N$ data instances, we denote the training data as $S=\left\{\left(X^{\mathrm{CT}}_n, X^{\mathrm{PET}}_n, Y_n\right)\right\}_{n=1}^{N}$, where $X^{\mathrm{CT}}_n$, $X^{\mathrm{PET}}_n$, and $Y_n$ represent the input \ac{CT}, registered \ac{PET}, and binary ground truth \ac{GTV} segmentation images, respectively. For simplicity and consistency, the same 3D segmentation backbone network (described in Sec.~\ref{sec:PSSN}) is adopted. Dropping $n$ for clarity, we first use two separate streams to generate segmentation maps using $X^{\mathrm{CT}}$ and  [$X^{\mathrm{CT}}$, $X^{\mathrm{PET}}$] as network input channels: 
%\hat{Y}^{\mathrm{CT}} = \left\{ \hat{y}^{\mathrm{CT}}_j\right\} 
%\hat{Y}^{\mathrm{EF}} = \left\{ \hat{y}^{\mathrm{EF}}_j\right\} \mathrm{,}
\begin{align}
\hat{y}^{\mathrm{CT}}_j&=p_j^{\mathrm{CT}}\left( y_j=1 | X^{\mathrm{CT}};  \mathbf{W}^{\mathrm{CT}}\right) \mathrm{,} \label{eqn:CT_stream} \\  
   \hat{y}^{\mathrm{EF}}_j &= p_j^{\mathrm{EF}}\left( y_j=1 | X^{\mathrm{CT}}, X^{\mathrm{PET}};  \mathbf{W}^{\mathrm{EF}} \right)  \mathrm{,} \label{eqn:EF}
\end{align}
where $p_j^{(\cdot)}(\cdot)$ and $\hat{y}^{(\cdot)}_j$ denote the \ac{CNN} functions and output segmentation maps, respectively, $\mathbf{W}^{(\cdot)}$ represents the corresponding \ac{CNN} parameters, and $y_j$ indicates the ground truth \ac{GTV} tumor mask values. We denote Eq. \eqref{eqn:EF} as \acf{EF}, as the stream can be seen as an \ac{EF} of \ac{CT} and \ac{PET}, enjoying the high spatial resolution and high tumor-intake contrast properties from the \ac{CT} and \ac{PET}, respectively. On the other hand, the stream in Eq. \eqref{eqn:CT_stream} provides predictions based on \ac{CT} intensity alone, which can be particularly helpful in circumventing the biased influence from noisy non-malignant high uptake regions, which are not uncommon in \ac{PET}.

As \Fig~\ref{fig:overall_pipeline}(a) illustrates, we harmonize the outputs from Eq. \eqref{eqn:CT_stream} and Eq. \eqref{eqn:EF} by concatenating them together with the original \ac{RTCT} image as the inputs to a third network:
\begin{align}
\hat{y}^{\mathrm{LF}}_j=p_j^{\mathrm{LF}}\left( y_j=1 | X^{\mathrm{CT}}, \hat{Y}^{\mathrm{CT}}, \hat{Y}^{\mathrm{EF}};  \mathbf{W}^{\mathrm{CT}},\mathbf{W}^{\mathrm{EF}},\mathbf{W}^{\mathrm{LF}}\right) \mathrm{.} \label{eqn:lf}
\end{align}
In this way, the formulation of Eq. \eqref{eqn:lf} can be seen as a \acf{LF} of the aforementioned two streams of the \ac{CT} and \ac{EF} models. We use the \ac{DSC} loss for all three sub-networks, training each in isolation. 

\subsection{\ac{PSNN} Model} \label{sec:PSSN}

In esophageal \ac{GTV} segmentation, the \ac{GTV} target region often exhibits low contrast in \ac{CT}, and the physician's manual delineation relies heavily upon high-level semantic information to disambiguate boundaries. In certain respects, this aligns with the intuition behind UNet, which decodes high-level features into lower-level space. Nonetheless, the decoding path in UNet consumes a great deal of parameters, adding to its complexity. On the other hand, models like P-HNN~\cite{harrison2017progressive} use deep supervision to connect lower and higher-level features together using parameter-less pathways. However, unlike UNet, P-HNN propagates lower-level features \emph{down} to high-level layers. Instead, a natural and simple means to combine the strengths of both P-HNN and UNet is to use essentially the same parameter blocks as P-HNN, but reverse the direction of the deeply-supervised pathways, to allow high-level information to propagate \emph{up} to lower-level space. We denote such an approach as \acf{PSNN}. 

As shown in \Fig~\ref{fig:overall_pipeline}(b), a set of $1\times 1 \times 1$ 3D convolutional layers are used to collapse the feature map after each convolutional block into a logit image, \ie{} $\tilde{f}_j^{(\ell)}$, where $j$ indexes the pixel locations. This is then combined with the previous higher level segmentation logit image to create an aggregated segmentation map, \ie{} $f_j^{(\ell)}$, for the $\ell^{th}$ feature block by element-wise summation:
\begin{align}
f_{j}^{(m)} & = \tilde{f}_{j}^{(m)} \textrm{,} \\
f_{j}^{(\ell)}&= \tilde{f}_{j}^{(\ell)} + g\left(f_{j}^{(\ell+1)}\right) , \forall \ell \in \{m-1, \cdots, 1\},
\label{eq:psa} 
\end{align}
where $m$ denotes the total number of predicted feature maps and $g(\cdot)$. denotes an upsampling, \ie{} bilinear upsampling. The \ac{PSNN} model is trained using four deeply-supervised auxiliary losses at each convolutional block. As our experiments will demonstrate, \ac{PSNN} can provide significant performance gains for \ac{GTV} segmentation over both a densely connected version of UNet~\cite{yousefi2018esophageal} and P-HNN~\cite{harrison2017progressive}.

\section{Experiments and Results}
\label{sec:experiment}

\begin{figure}[t]
\centering
\includegraphics[width=0.93\columnwidth]{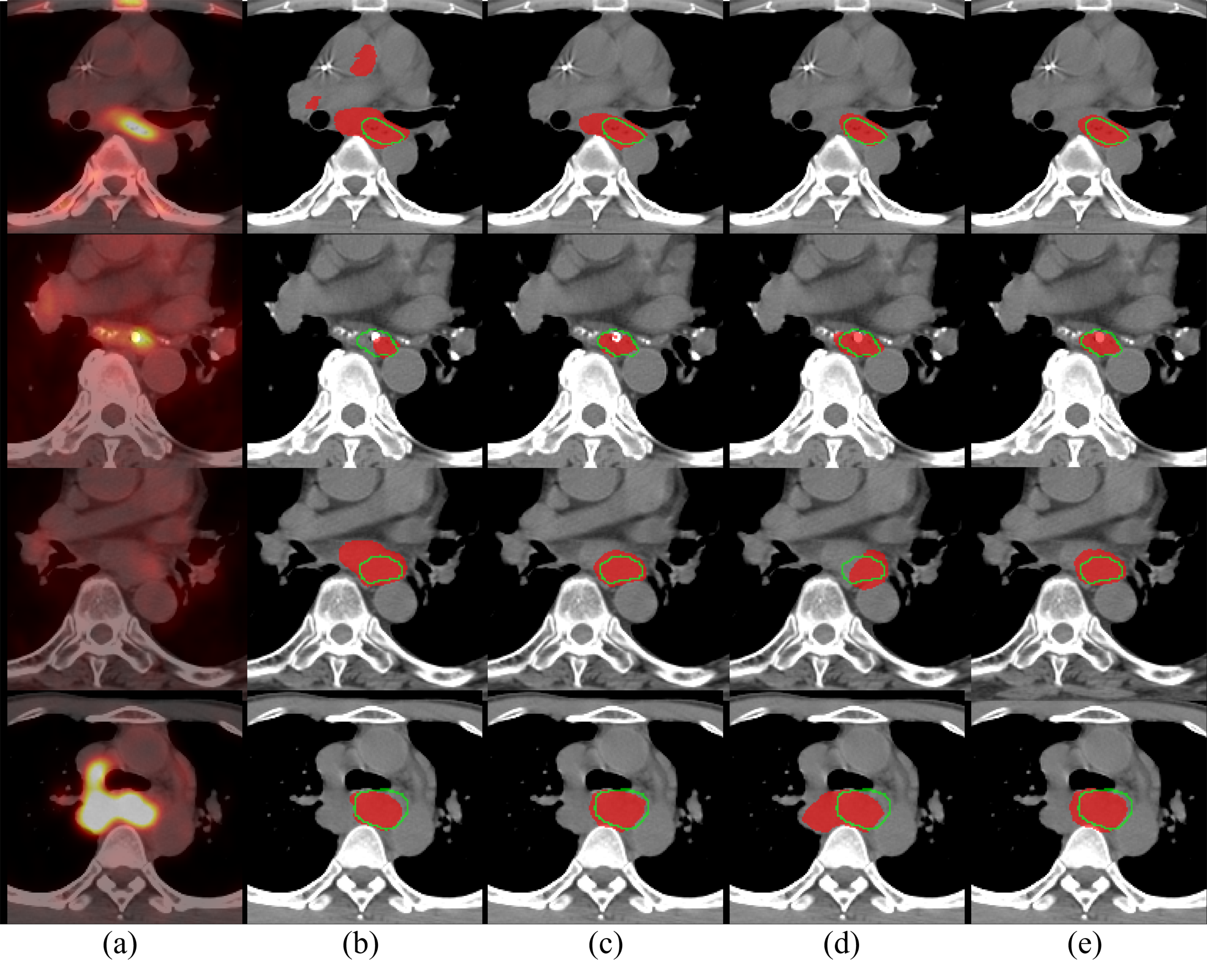}
\caption{Qualitative results of esophageal GTV segmentation. (a) RTCT overlayed with the registered PET channel; (b) GTV segmentation results using RTCT images with DenseUNet~\cite{yousefi2018esophageal}; (c) Our results using Eq. \eqref{eqn:CT_stream}, \ie{} RTCT-only stream using the proposed \ac{PSNN} model; (d) Our results using Eq. \eqref{eqn:EF} with \ac{PSNN}, \ie{} \acs{EF} of PET+RTCT images; (e) Our final GTV segmentation using Eq. \eqref{eqn:lf} with \ac{PSNN}, \ie{} \acs{EF} and \acs{LF} of PET+RTCT images. Red masks indicate automated segmentation results and green boundaries represent the ground truth. The first two rows demonstrate the importance of PET as using RTCT alone can cause under- or over-segmentation due to low contrast. The last two rows show cases where under- or over-segmentation can occur when the PET imaging channel is spuriously noisy. In all cases, the final \ac{EF}+\ac{LF} based GTV segmentation results achieve good accuracy and robustness.}
\label{fig:gan_quality} \vspace{-4mm}
\end{figure}

We extensively evaluate our approach using a dataset of $110$ esophageal cancer patients, all diagnosed at stage II or later and undergoing \ac{RT} treatments. Each patient has a diagnostic \ac{PET/CT} pair and a treatment \ac{RTCT} scan. To the best of our knowledge, this is the largest dataset collected for esophageal cancer \ac{GTV} segmentation. All 3D \ac{GTV} ground truth masks are delineated by two experienced radiation oncologists during routine clinical workflow. We first resample all imaging scans of registered \ac{PET} and \ac{RTCT} to a fixed resolution of $1.0\times1.0\times2.5$ mm. To generate positive training instances, we randomly sample $80 \times 80 \times 64$ sub-volumes centered inside the ground truth \ac{GTV} mask. Negative examples are extracted by randomly sampling from the whole 3D volume. This results, on average, in $80$ training sub-volumes per patient. We further apply random rotations in the x-y plane within $\pm10$ degrees to augment the training data.

\begin{table}[t]
\centering
\begin{small}
\caption{Mean DSCs, HDs, and $\text{ASD}_{GT}$, and their standard deviations of \ac{GTV} segmentation  performance using: (1) Contextual model using only CT images (CT); (2) Early fusion model (EF) using both CT and PET images; (3) The proposed two-stream chained early and late fusion model (EF+LF). 3D DenseUNet model using CT is equivalent to the previous state-of-the-art work~\cite{yousefi2018esophageal}, which is shown in the first row. The best performance scores are shown in {\bf bold}.} \label{tbl:results}
\begin{tabular}{|l|ccc|ccc|}
\hline
                           & CT         & EF     & EF+LF         & DSC  & HD (mm)          & $\text{ASD}_{GT}$ (mm) \\ \hline
\multirow{3}{*}{3D DenseUNet} & \checkmark &            &             & 0.654$\pm$0.210 & 129.0$\pm$73.0 & 5.2$\pm$12.8 \\
                           &            & \checkmark &             & 0.710$\pm$0.189 & 116.0$\pm$81.7 & 4.9$\pm$10.3 \\
                           &            &            & \checkmark  & 0.745$\pm$0.163 & 79.5$\pm$70.9 & 4.7$\pm$10.5 \\ \hline
\multirow{3}{*}{3D P-HNN} & \checkmark &            &             & 0.710$\pm$0.189 & 86.2$\pm$67.4 & 4.3$\pm$5.3 \\
                           &            & \checkmark &             & 0.735$\pm$0.158 & 57.9$\pm$61.1 & 3.6$\pm$3.7 \\
                           &            &            & \checkmark  & 0.755$\pm$0.148 & \textbf{47.2$\pm$52.3} & 3.8$\pm$4.8 \\ \hline
\multirow{3}{*}{3D PSNN}   & \checkmark &            &             & 0.728$\pm$0.158 & 66.9$\pm$59.2 & 4.2$\pm$5.4 \\
                           &            & \checkmark &             & 0.758$\pm$0.136 & 67.0$\pm$59.1 & \textbf{3.2$\pm$3.1}  \\
                           &            &            & \checkmark  & \textbf{0.764$\pm$0.134} & \textbf{47.1$\pm$56.0} & \textbf{3.2$\pm$3.3}  \\ \hline
\end{tabular}
\end{small}
\end{table}

\noindent\textbf{Implementation details:} The Adam solver~\cite{kingma2014adam} is used to optimize all the 3D segmentation models with a momentum of $0.99$ and a weight decay of $0.005$ for $40$ epochs. For testing, we use 3D sliding windows with sub-volumes of $80 \times 80 \times 64$ and strides of $48 \times 48 \times 32$ voxels. The probability maps of sub-volumes are aggregated to obtain the whole volume prediction. %taking \jin{on average $10\,\mathrm{s}$} to process one input volume using a single Titan-V GPU.

We employ five-fold cross-validation protocol split at the patient level. Extensive comparisons of our \ac{PSNN} model versus P-HNN~\cite{harrison2017progressive} and DenseUNet~\cite{yousefi2018esophageal} methods are reported, with the latter arguably representing the current state-of-the-art \ac{GTV} segmentation approach using CT. Three quantitative metrics are utilized to evaluate the \ac{GTV} segmentation performance: \ac{DSC}, \ac{HD} in ``mm", and \ac{ASD} in ``mm".

\noindent\textbf{Results:} Our quantitative results and comparisons are tabulated in Table~\ref{tbl:results}. When all models are trained and evaluated using only \ac{RTCT}, \ie{} Eq. \eqref{eqn:CT_stream}, our proposed \ac{PSNN} evidently outperforms the previous best esophageal GTV segmentation method, \ie{} DenseUNet~\cite{yousefi2018esophageal}, which straightforwardly combines DenseNet~\cite{DBLP:conf/cvpr/HuangLMW17} and 3D UNet~\cite{Cicek2016}. As can be seen, \ac{PSNN} consistently improves upon~\cite{yousefi2018esophageal} in all metrics: with an absolute increase of $7.4\%$ in \ac{DSC} (from $0.654$ to $0.728$) and significantly dropping in \ac{HD} metric, despite being a simpler architecture.  \ac{PSNN} also outperforms the 3D version of P-HNN~\cite{harrison2017progressive}, which indicates that the semantically-nested high- to low-level information flow provides key performance increases. 

Table~\ref{tbl:results} also outlines the performances of three deep models under different imaging configurations. Several conclusions can be drawn. First, all three networks trained using the \ac{EF} of Eq. \eqref{eqn:EF} consistently produce more accurate segmentation results than those trained with only \ac{RTCT}, \ie{} Eq. \eqref{eqn:CT_stream}. This validates the effectiveness of utilizing \ac{PET} to complement \ac{RTCT} for \ac{GTV} segmentation. Second, the full two-stream chained fusion pipeline of Eq. \eqref{eqn:lf} provides further performance improvements. Importantly, the performance boosts can be observed across all three deep \acp{CNN}, validating that the two-stream combination of \ac{EF} and \ac{LF} can universally improve upon different backbone segmentation models. Last, the best performing results are the \ac{PSNN} model combined with chained \ac{EF}+\ac{LF}, demonstrating that each component of the system contributes to our final performance. When compared to the previous state-of-the-art work of \ac{GTV} segmentation, which uses DenseUNet applied to \ac{RTCT} images~\cite{yousefi2018esophageal}, our best performing model exceeds in all metrics of \ac{DSC}, \ac{HD}, and \ac{ASD} by $11\%$, $81.9mm$ and $2.0mm$ remarked margins (refer to Table~\ref{tbl:results}), representing tangible and significant improvements. \Fig~\ref{fig:gan_quality} shows several qualitative examples visually underscoring the improvements that our two-stage \ac{PSNN} approach provides.

\section{Conclusion}
\label{sec:conclusion}

This work has presented and validated a two-stream chained 3D deep network fusion pipeline to segment esophageal \acp{GTV} using both RTCT and PET+RTCT imaging channels. Diagnostic PET and RTCT are first longitudinally registered using semantically-based lung-mass center initialization to achieve robustness. We next employ the \ac{PSNN} model as a new 3D segmentation architecture, which uses a simple, parameter-less, and deeply-supervised CNN decoding stream. The \ac{PSNN} model is then used in a cascaded \ac{EF} and \ac{LF} scheme to segment the \ac{GTV}. Extensive tests on the largest esophageal dataset to date demonstrate that our \ac{PSNN} model can outperform the state-of-the-art P-HNN and DenseUNet networks with remarked margins. Additionally, we show that our 2-stream chained fusion pipeline produces further important improvements, providing an effective means to exploit the complementary information seen within PET and CT. Thus, our work represents a step forward toward accurate and automated esophageal \ac{GTV} segmentation.

\bibliographystyle{splncs}
\small{\bibliography{refs}}
\end{document}